\documentclass[twocolumn]{aastex6}

\usepackage{amsfonts, amsmath, amsthm, amssymb}

\newcommand{\mbf}[1]{\mbox{\boldmath $#1$}}
\newcommand{\mbfs}[1]{\mbox{\scriptsize\boldmath $#1$}}

\newcommand{\Eqn}[1]{Equation~(\ref{eqn:#1})}
\newcommand{\Eqns}[1]{Equations~(\ref{eqn:#1})}
\newcommand{\eqn}[1]{equation~(\ref{eqn:#1})}
\newcommand{\eqns}[1]{equations~(\ref{eqn:#1})}

\newcommand{\Sec}[1]{\S~\ref{sec:#1}}

\newcommand{\Fig}[1]{Figure~\ref{fig:#1}}
\newcommand{\Figs}[2]{Figures~\ref{fig:#1} and~\ref{fig:#2}}

\newcommand{\App}[1]{Appendix~\ref{app:#1}}

\newcommand{\Ci}{\ensuremath{i}}

\newcommand{\trace}{{\rm Tr}}

\newcommand{\Rotation}{{\bf R}}
\newcommand{\Boost}{{\bf B}}

\newcommand{\vRotation}[1][n]{\ensuremath{\Rotation_{\mbfs{\hat #1}}}}
\newcommand{\vBoost}[1][m]{\ensuremath{\Boost_{\mbfs{\hat #1}}}}

\newcommand{\mean}[1][ ]{\ensuremath{ \hat{#1} }}
\newcommand{\smean}[1][ ]{\ensuremath{ \bar{#1} }}
\newcommand{\inner}[2]{\ensuremath{ {\langle #1, #2 \rangle} }}

\newcommand{\rotat}{\ensuremath{\vRotation(\phi)}}
\newcommand{\boost}{\ensuremath{\vBoost(\beta)}}

\newcommand{\pauli}[1]{\ensuremath{\mbf{\sigma}_{#1}}}

\newcommand{\inv}[1]{\ensuremath{ {#1}^2 }}
\newcommand{\norm}[1]{\ensuremath{ \|{#1}\| }}

\newcommand{\acomm}[2]{\ensuremath{ \left\{{#1},{#2}\right\} }}

\shorttitle   {Statistics of Polarimetry}
\shortauthors {W. van Straten}

\begin{document}

\title{ The Statistics of Radio Astronomical Polarimetry: \\
	Bright Sources and High Time Resolution }

\author{W. van Straten}

\affil{Centre for Astrophysics and Supercomputing,
	Swinburne University of Technology, \\
	Hawthorn, VIC 3122, Australia}

\email{vanstraten.willem@gmail.com}

\begin{abstract}

A four-dimensional statistical description of electromagnetic
radiation is developed and applied to the analysis of radio pulsar
polarization.  The new formalism provides an elementary statistical
explanation of the modal broadening phenomenon in single pulse
observations.  It is also used to argue that the degree of
polarization of giant pulses has been poorly defined in past studies.
Single and giant pulse polarimetry typically involves sources with
large flux densities and observations with high time resolution,
factors that necessitate consideration of source-intrinsic noise and
small-number statistics.  Self noise is shown to fully explain the
excess polarization dispersion previously noted in single pulse
observations of bright pulsars, obviating the need for additional
randomly polarized radiation.  Rather, these observations are more
simply interpreted as an incoherent sum of covariant, orthogonal,
partially polarized modes.  Based on this premise, the
four-dimensional covariance matrix of the Stokes parameters may be
used to derive mode-separated pulse profiles without any assumptions
about the intrinsic degrees of mode polarization.  Finally, utilizing
the small-number statistics of the Stokes parameters, it is
established that the degree of polarization of an unresolved pulse is
fundamentally undefined; therefore, previous claims of highly
polarized giant pulses are unsubstantiated.
Unpublished supplementary material is appended after the bibliography.

\end{abstract}


%%%%%%%%%%%%%%%%%%%%%%%%%%%%%%%%%%%%%%%%%%%%%%%%%%%%%%%%%%%%%%%%%%%%%%
%%%%%%%%%%%%%%%%%%%%%%%%%%%%%%%%%%%%%%%%%%%%%%%%%%%%%%%%%%%%%%%%%%%%%%
%%%%%%%%%%%%%%%%%%%%%%%%%%%%%%%%%%%%%%%%%%%%%%%%%%%%%%%%%%%%%%%%%%%%%%

\section {Introduction}

Radio pulsars exhibit dramatic fluctuations in total and polarized flux
densities on a diverse range of longitudinal, temporal and spectral
scales.
As a function of pulse phase, both average profiles and sub-pulses
make sudden transitions between orthogonally polarized modes
\citep[e.g.][]{thhm71,mth75}.
The radiation at a single pulse phase often appears as an incoherent
superposition of modes, both orthogonal and non-orthogonal
\citep[e.g.][]{br80,mck03a}.
Evidence has also been presented for variations that may indicate
stochastic generalized Faraday rotation in the pulsar magnetosphere
\citep{es04}.

In an extensive study of single-pulse polarization fluctuations at
1400~MHz, \citet{scr+84} observed that histograms of the linear
polarization position angle were broader than could be explained
by instrumental noise alone.
This modal broadening of the position angle distribution was
interpreted as evidence of a superposed randomized emission component.
\citet{ms98} revisited the issue with a statistical model that 
described correlated intensity fluctuations of completely polarized
orthogonal modes.
Although the predicted position angle distributions were qualitatively
similar to the observations, the measured histograms were wider than
those produced by a numerical simulation of modal broadening that
included instrumental noise.
Source-intrinsic noise was later used to explain the excess
polarization scatter of bright pulses in an analysis of mode-separated
profiles \citep{ms00}; however, it was given no further consideration
in all subsequent statistical treatments by
\citet{mck02,mck03b,mck04,mck06}.

These begin with the reasonable assumptions that the instantaneous
signal-to-noise ratio $S/N$ is low and a sufficiently large number of
samples have been averaged, such that the stochastic noise in all four
Stokes parameters can be treated as uncorrelated and normally
distributed.
These assumptions form part of the three-dimensional eigenvalue
analyses of the Stokes polarization vector by \citet{mck04} and
\citet{es04}.
As in \cite{scr+84}, these studies concluded that modal
broadening is due to the incoherent addition of randomly polarized
radiation that is intrinsic to the pulsar.
The basic premises of these experiments are valid for the vast
majority of pulsar observations; however, they become untenable for
the bright sources on which these studies focus.
That is, when the instantaneous $S/N \gtrsim 1$, the Stokes
polarization vector can no longer be treated in isolation and the
correlated self noise intrinsic to all four stochastic Stokes
parameters must be accounted.

These considerations are particularly relevant to the study of giant
pulses.  Those from the Crab pulsar reach brightness temperatures in
excess of $10^{37}$ Kelvin and remain unresolved in observations with
nanosecond resolution \citep{hkwe03}.
Consequently, previous analyses of giant pulses have typically
presented polarization data at the sampling resolution of the
instrument; that is, where the time-bandwidth product is of the order
of unity.
For example, \citet{hcr70} studied giant pulses from the Crab pulsar
using an instrument with bandwidth, $\delta\nu=8.3$~kHz, and presented
plots of the Stokes parameters at a resolution of $\tau=120\mu$s.
\citet{cstt96} plotted the total intensity and circular polarization
of the strongest giant pulse and interpulse observed from PSR~B1937+21
with $\tau=1.2\mu$s and $\delta\nu=500$~kHz.
Using a baseband recorder with $\delta\nu=500$~MHz, \citet{hkwe03}
plotted the intensities of left and right circularly polarized
radiation from Crab giant pulses at a time resolution of 2~ns.
Each of these studies concluded that giant pulses are highly
polarized.
However, in each experiment, $\tau\delta\nu\sim1$; at this resolution,
every discrete sample of the electric field is completely polarized,
regardless of the intrinsic degree of polarization of the source.
That is, the instantaneous degree of polarization is fundamentally
undefined.

To study the polarization of intense sources of impulsive radiation at
high time resolution, it is necessary to consider averages over small
numbers of samples and source-intrinsic noise statistics.
These limitations are given careful attention in the statistical
theory of polarized shot noise \citep{cor76}.
This seminal work enables characterization of the timescales of
microstructure polarization fluctuations via the auto-correlation
functions of the total intensity and degrees of linear and circular
polarization \citep[e.g.][]{ch77}.
It has also been extended to study the degree of polarization via the
cross-correlation as a function of time lag between the instantaneous
total intensity spectra of giant pulses \citep{cbh+04}.

This paper presents a complimentary approach based on the
four-dimensional joint distribution and covariance matrix of the
Stokes parameters, with emphasis placed on the statistical degrees of
freedom of the underlying stochastic process.
Relevant theoretical results are drawn from various sources, ranging
from studies of the scattering of monochromatic light in optical fibres
\citep[e.g.][]{sazf84} to the classification of synthetic aperture
radar images \citep[e.g.][]{tl96}.

Following a brief review of polarization algebra in \Sec{review}, the
joint probability density functions of the Stokes parameters at
different resolutions are derived in \Sec{joint}, where the results
are compared and contrasted with previous works.
The formalism is related to the study of radio pulsar polarization in
\Sec{application}, where the effects of amplitude modulation 
and wave coherence on the degrees of freedom of the pulsar
signal are discussed.
Finally, the results are utilized to reexamine past statistical analyses
of orthogonally polarized modes, randomly polarized radiation, and
giant pulse polarimetry.
It is concluded that randomly polarized radiation is unnecessary and
the degree of polarization of giant pulses must be more rigorously
defined.
Potential applications of the four-dimensional statistics of polarized
noise are proposed in \Sec{conclusion}.

%%%%%%%%%%%%%%%%%%%%%%%%%%%%%%%%%%%%%%%%%%%%%%%%%%%%%%%%%%%%%%%%%%%%%%
%%%%%%%%%%%%%%%%%%%%%%%%%%%%%%%%%%%%%%%%%%%%%%%%%%%%%%%%%%%%%%%%%%%%%%
%%%%%%%%%%%%%%%%%%%%%%%%%%%%%%%%%%%%%%%%%%%%%%%%%%%%%%%%%%%%%%%%%%%%%%

\section{Polarization Algebra}
\label{sec:review}

This section reviews the relevant algebra of both Jones and Mueller
representations of polarimetric transformations.  Unless otherwise
noted, the notation and terminology synthesizes that of the similar
approaches to polarization algebra presented by \citet{clo86},
\citet{bri00}, and \cite{ham00}.

\subsection{Jones Calculus}
\label{sec:jones}

The polarization of electromagnetic radiation is described by the
second-order statistics of the transverse electric field vector,
$\mbf{e}$, as represented by the complex $2\times2$ coherency matrix,
$\mbf{\rho}\equiv\langle\mbf{e\, e}^\dagger\rangle$ \citep{bw80}.  Here,
the angular brackets denote an ensemble average, $\mbf{e}^\dagger$ is
the Hermitian transpose of $\mbf{e}$, and an outer product is implied
by the treatment of $\mbf{e}$ as a column vector.  A useful geometric
relationship between the complex two-dimensional space of the
coherency matrix and the real four-dimensional space of the Stokes
parameters is expressed by the following pair of equations:
\begin{eqnarray}
{\mbf\rho} & = & S_k\,\pauli{k} / 2     \label{eqn:combination} \\
S_k & = & \trace(\pauli{k}\mbf{\rho}).  \label{eqn:projection}
\end{eqnarray}
Here, $S_k$ are the four Stokes parameters, Einstein notation is used
to imply a sum over repeated indeces, $0\le k \le3$, $\pauli{0}$ is
the $2\times2$ identity matrix, $\pauli{1-3}$ are the Pauli matrices,
and $\trace$ is the matrix trace operator.  The Stokes four-vector is
composed of the the total intensity $S_0$ and the polarization vector,
$\mbf{S}=(S_1,S_2,S_3)$.  \Eqn{combination} expresses the coherency
matrix as a linear combination of Hermitian basis matrices;
\eqn{projection} represents the Stokes parameters as the projections
of the coherency matrix onto the basis matrices.  
These properties are used to interpret the well-studied statistics of
random matrices through the familiar Stokes parameters.

Linear transformations of the electric field vector are represented
using complex $2\times2$ Jones matrices.  Substitution of
$\mbf{e}^\prime={\bf J}\mbf{e}$ into the definition of the coherency
matrix yields the congruence transformation,
\begin{equation}
{\mbf{\rho}^\prime}={\bf{J}}\mbf{\rho}{\bf{J}}^\dagger,
\label{eqn:congruence}
\end{equation}
%%
%% REF 4.
%%
which forms the basis of the various coordinate transformations that
are exploited throughout this work.
If ${\bf{J}}$ is non-singular, it can be decomposed into the product
of a Hermitian matrix and a unitary matrix known as its polar
decomposition,
\begin{equation}
\label{eqn:polar}
{\bf J} = J \, \boost \, \rotat,
\end{equation}
where $J=|{\bf J}|^{1/2}$, \boost\ is positive-definite Hermitian, and
\rotat\ is unitary; both \boost\ and \rotat\ are unimodular.  Under
the congruence transformation of the coherency matrix, the Hermitian
matrix
\begin{equation}
\label{eqn:Boost}
\boost = \exp (\beta\,\mbf{\hat{m}\cdot\sigma})
       = \pauli{0}\cosh\beta + \mbf{\hat{m}\cdot\sigma}\sinh\beta
\end{equation}
effects a Lorentz boost of the Stokes four-vector along the $\mbf{\hat m}$
axis by a hyperbolic angle $2\beta$.
%%
%% REF 4.
%%
As the Lorentz transformation of a spacetime event mixes temporal and
spatial dimensions, the polarimetric boost mixes total and polarized
intensities, thereby altering the degree of polarization.
In contrast, the unitary matrix
\begin{equation}
\label{eqn:Rotation}
\rotat = \exp (\Ci\phi\,\mbf{\hat{n}\cdot\sigma})
       = \pauli{0}\cos\phi + \Ci\mbf{\hat{n}\cdot\sigma}\sin\phi
\end{equation}
rotates the Stokes polarization vector about the $\mbf{\hat n}$ axis by
an angle $2\phi$.
%%
%% REF 4.
%%
As the orthogonal transformation of a vector in Euclidean space
preserves its length, the polarimetric rotation leaves the degree of
polarization unchanged.

These geometric interpretations promote a more intuitive treatment of
the matrix equations that typically arise in polarimetry.
Boost transformations can be utilized to convert unpolarized radiation
into partially polarized radiation, and rotation transformations can
be used to choose the orthonormal basis that maximizes symmetry.
These properties are exploited in \Sec{joint} to simplify the relevant
mathematical expressions that describe the four-dimensional joint
distribution of the Stokes parameters.

%%%%%%%%%%%%%%%%%%%%%%%%%%%%%%%%%%%%%%%%%%%%%%%%%%%%%%%%%%%%%%%%%%%%%%
%%%%%%%%%%%%%%%%%%%%%%%%%%%%%%%%%%%%%%%%%%%%%%%%%%%%%%%%%%%%%%%%%%%%%%

\subsection{Eigen Decomposition}
\label{sec:eigen}

It proves useful in \Sec{joint} to express the coherency matrix
as a similarity transformation known as its eigen decomposition,
\begin{equation}
\mbf{\rho}={\bf{R}}\left( \begin{array}{cc}
\lambda_0 & 0 \\
0 & \lambda_1
\end{array}\right){\bf{R}}^{-1}.
\label{eqn:eigen}
\end{equation}
Here, ${\bf{R}}=(\mbf{e}_0\,\mbf{e}_1)$ is a $2\times2$ matrix with
columns equal to the eigenvectors of $\mbf{\rho}$, and $\lambda_m$ are
the corresponding eigenvalues, given by
$\lambda=(S_0\pm|\mbf{S}|)/2=(1 \pm P)S_0/2$, where $P=S_0/|\mbf{S}|$
is the degree of polarization.  If the signal is completely
polarized, then $\lambda_1=0$.  If the signal is unpolarized, then
there is a single 2-fold degenerate eigenvalue, $\lambda=S_0/2$ and
${\bf R}$ is undefined.

If the eigenvectors are normalized such that
${\mbf{e}_k}^\dagger\mbf{e}_k=1$, then \eqn{eigen} is equivalent to a
congruence transformation by the unitary matrix, ${\bf{R}}$.
In the natural basis defined by ${\bf{R}}^\dagger$, the eigenvalues
$\lambda_m$ are equal to the variances of two uncorrelated signals
received by orthogonally polarized receptors described by the
eigenvectors.
The total intensity, $S_0=\lambda_0+\lambda_1$; the
polarized intensity, $S_1=|\mbf{S}|=\lambda_0-\lambda_1$; and
$S_2=S_3=0$.  
That is, ${\bf{R}}^\dagger$ rotates the basis such that the mean
polarization vector points along $S_1$, providing cylindrical symmetry
about this axis.

%%%%%%%%%%%%%%%%%%%%%%%%%%%%%%%%%%%%%%%%%%%%%%%%%%%%%%%%%%%%%%%%%%%%%%
%%%%%%%%%%%%%%%%%%%%%%%%%%%%%%%%%%%%%%%%%%%%%%%%%%%%%%%%%%%%%%%%%%%%%%

\subsection{Mueller Calculus}
\label{sec:mueller}

The congruence transformation of the coherency matrix by any Jones
matrix {\bf J} may be represented by an equivalent linear
transformation of the Stokes parameters by a real-valued $4\times4$
Mueller matrix
\begin{equation}
\label{eqn:Mueller}
M_i^k = \frac{1}{2}\trace(\pauli{i}\,{\bf J}\,\pauli{k}\,{\bf J}^\dagger),
\end{equation}
such that
\begin{equation}
S^\prime_i = M_i^k S_k.
\end{equation}
Mueller matrices that have an equivalent Jones matrix are called pure,
and such transformations are related to the Lorentz group
\citep[e.g.][]{bar63,bri00}.  
This motivates the definition of an inner product
\begin{equation}
\inner{A}{B} \equiv A^k B_k = \eta^{kk}A_k B_k = A_0B_0 - \mbf{A \cdot B},
\end{equation}
where $A$ and $B$ are Stokes four-vectors and $\eta_{ij}$ is the
Minkowski metric tensor with signature $(+,-,-,-)$.
The Lorentz invariant of a Stokes four-vector is equal to four times
the determinant of the coherency matrix; that is,
\begin{equation}
\inv{S} \equiv \inner{S}{S} = S_0^2 - |\mbf{S}|^2 = 4 |\mbf{\rho}|
\end{equation}
%%
%% REF 21
%%
Similarly, the Euclidean norm \norm{S} is twice
the Frobenius norm of the coherency matrix \norm{\mbf{\rho}}; i.e.
\begin{equation}
\norm{S}^2 \equiv S_0^2 + |\mbf{S}|^2 = 4 \norm{\mbf{\rho}}^2.
\end{equation}
The coherency matrix is a positive semi-definite Hermitian matrix;
therefore, the Lorentz invariant of any physically realizable source
of radiation is greater than or equal to zero.  It is equal to zero
only for completely polarized radiation, when the degree of
polarization is unity ($S_0=|\mbf{S}|$).  As with the spacetime null
interval, no linear transformation of the electric field can alter the
degree of polarization of a completely polarized source.

%%%%%%%%%%%%%%%%%%%%%%%%%%%%%%%%%%%%%%%%%%%%%%%%%%%%%%%%%%%%%%%%%%%%%%
%%%%%%%%%%%%%%%%%%%%%%%%%%%%%%%%%%%%%%%%%%%%%%%%%%%%%%%%%%%%%%%%%%%%%%
%%%%%%%%%%%%%%%%%%%%%%%%%%%%%%%%%%%%%%%%%%%%%%%%%%%%%%%%%%%%%%%%%%%%%%

\section{Joint Distribution Function}
\label{sec:joint}

In this section, the joint distribution functions of the Stokes
parameters for a stationary stochastic source of polarized radiation
are derived.  Three regimes of interest are considered: single
samples, local means, and ensemble averages of large numbers of
samples.
The electric field vector is assumed to have a jointly normal density
function as described by \cite{goo63}.
This distribution may not accurately describe the possibly non-linear
electric field of intense non-thermal radiation.
For example, over a limited range of pulse phase, the fluctuations in
the intensity of the Vela pulsar have a lognormal distribution that is
consistent with the predictions of stochastic growth theory
\citep{cjd01}.
The normal distribution is a valid choice if the pulsar signal can be
accurately modeled as polarized shot-noise, provided that the density
of shots is sufficiently high \citep{cor76}.
It is also the minimal assumption if no information about the
higher-order moments of the field is available.

%%%%%%%%%%%%%%%%%%%%%%%%%%%%%%%%%%%%%%%%%%%%%%%%%%%%%%%%%%%%%%%%%%%%%%
%%%%%%%%%%%%%%%%%%%%%%%%%%%%%%%%%%%%%%%%%%%%%%%%%%%%%%%%%%%%%%%%%%%%%%

\subsection{Single Samples}
\label{sec:single}

As second-order moments of the electric field, the coherency matrix
and Stokes parameters are defined only after an average is made over
some number of samples.  
Given a single instance of the electric field $\mbf{e}$ (for example,
discretely sampled at an infinitesimal moment in time) define the
instantaneous coherency matrix, ${\bf{r}}=\mbf{e\,e}^\dagger$, and
Stokes parameters, $s_k=\trace(\pauli{k}{\bf{r}})$, such that the
ensemble averages, $\mbf{\rho}=\langle{\bf{r}}\rangle$ and
$S_k=\langle{s_k}\rangle$.
It is trivial to show that the determinant of the instantaneous
coherency matrix, $|\mbf{r}|=0$, regardless of the degree of
polarization of the source or the probability density of the electric
field.
That is, the instantaneous degree of polarization is undefined.

The complex-valued components of $\mbf{e}$ are the analytic signals
associated with the real-valued voltages measured in each receptor.
If the voltages are normally distributed with zero mean, then $\mbf e$
has a bivariate complex normal distribution
\citep{goo63},
\begin{equation}
f(\mbf{e}) = {1\over\pi^2|\mbf\rho|}
\exp \left( - \mbf{e}^\dagger \mbf{\rho}^{-1} \mbf{e} \right),
\end{equation}
where $\mbf\rho=S_k\pauli{k}/2$ is the population mean coherency
matrix.  To compute the probability density function of the
instantaneous Stokes parameters, note that $f(\mbf{e})$ is independent
of the absolute phase of $\mbf{e}$.  Conversion to polar coordinates
and marginalization over this variable yields the intermediate result,
\begin{equation}
f(|e_0|,|e_1|,\psi) = {2|e_0||e_1|\over\pi|\mbf\rho|}
\exp \left( - \mbf{e}^\dagger \mbf{\rho}^{-1} \mbf{e} \right),
\label{eqn:single_modes}
\end{equation}
where $\psi$ is the instantaneous phase difference between $e_0$ and
$e_1$.  This result may be compared with equation (2.10) of
\citet{bar87}, except that here the evaluation of the inner product
has been postponed.  The three remaining degrees of freedom are
described by the instantaneous polarization vector $\mbf{s}$, for
which the Jacobian determinant is
\begin{equation}
J({\mbf{s}}; |e_0|,|e_1|,\psi)
 = \left|{\partial\mbf{s}\over\partial(|e_0|,|e_1|,\psi)}\right| 
 = 8 |e_0||e_1||\mbf{s}|.
\end{equation}
Application of the above with
\begin{equation}
\label{eqn:inverse}
{\mbf\rho}^{-1}={1\over2|\mbf{\rho}|}(S_0\,\pauli{0}-\mbf{S\cdot\sigma})
= {2\over S^2} S^k \pauli{k}
\end{equation}
and
\begin{equation}
\mbf{e}^\dagger \mbf{\rho}^{-1} \mbf{e}
 = \trace \left( \mbf{e}^\dagger \mbf{\rho}^{-1} \mbf{e} \right)
 = \trace \left( \mbf{\rho}^{-1} \mbf{r} \right)
\end{equation}
yields the four-dimensional joint distribution of the instantaneous
Stokes parameters,
\begin{equation}
f(s) = { \delta(s_0-|\mbf{s}|) \over\pi \inv{S} s_0}
\exp \left( - { 2 \inner{S}{s} \over \inv{S} }  \right),
\label{eqn:single}
\end{equation}
where $\delta$ is the Dirac delta function and $s_0$ is the
instantaneous total intensity. 
This result is consistent with equation (11) of \citet{eli94}.
Converting to spherical polar coordinates, the inner product
$\inner{S}{s}=s_0(S_0-|\mbf{S}|)\cos\theta$, where $\theta$ is the
angle between $\mbf{s}$ and $\mbf{S}$.  Subsequent integration over
$\mbf{s}$ yields the marginal distribution of the instantaneous total
intensity,
\begin{equation}
f(s_0)={2\over|\mbf{S}|} 
       \exp\left(-{2 s_0 S_0 \over S^2}\right)
       \sinh\left({2 s_0 |\mbf{S}| \over S^2}\right),
\label{eqn:single_intensity}
%
% derivation in single_spherical.nb
%
\end{equation}
which has mean $S_0$ and variance $\norm{S}^2/2$.  This distribution is
consistent with equation (13) of \cite{man63} and equation (4) of
\cite{fs81}.  As noted by these authors, the total intensity becomes
$\chi^2$ distributed in the two cases of unpolarized and completely
polarized radiation.
This is easily seen in the natural basis defined by the eigen
decomposition of the coherency matrix.
As the squared norm of a complex number, the instantaneous intensity
in each of the two orthogonally polarized modes is $\chi^2$
distributed with two degrees of freedom.
If the signal is unpolarized, then each mode contributes identically
and the total intensity is $\chi^2$ distributed with four degrees of
freedom.
If the signal is 100\% polarized, then only one mode contributes and
the total intensity is $\chi^2$ distributed with two degrees of
freedom.

The marginal distributions of the instantaneous Stokes polarization
vector components are most easily derived in the natural basis, where
$\inner{S}{s}=s_0S_0-s_1S_1$.
Converting \mbf{s} to cylindrical coordinates with the axis of
symmetry along $S_1$ then integrating \eqn{single} over the radial
and azimuthal dimensions (as well as the total intensity) yields the
marginal distribution of the instantaneous major polarization,
\begin{equation}
f(s_1)={1\over S_0} 
       \exp\left( -{ 2 |s_1| S_0     \over S^2 } \right)
       \exp\left(  { 2 s_1 |\mbf{S}| \over S^2 } \right).
\label{eqn:single_major}
%
% derivation in single_cylindrical.nb
%
\end{equation}
This is an asymmetric Laplace density with mean $S_1$ and variance
$\norm{S}^2/2$, consistent with equation (8) of \cite{fs81}.
% Note the small error in Equation 8, 
% where \alpha_x\alpha_y should be (\alpha_x + \alpha_y)
The marginal distribution of the instantaneous minor polarization $s_2$
is derived in \App{marginal_minor}; by symmetry,
\eqn{marginal_minor} yields
\begin{equation}
f(s_{2,3})={1\over S} \exp\left( -{ 2 |s_{2,3}| \over S } \right),
\label{eqn:single_minor}
%
% derivation in single_eigen.nb
%
\end{equation}
where $S$ is the Lorentz interval.  This is a symmetric Laplace
density with mean zero and variance $\inv{S}/2$, as in equation (16)
of \cite{fs81}.

\begin{figure}
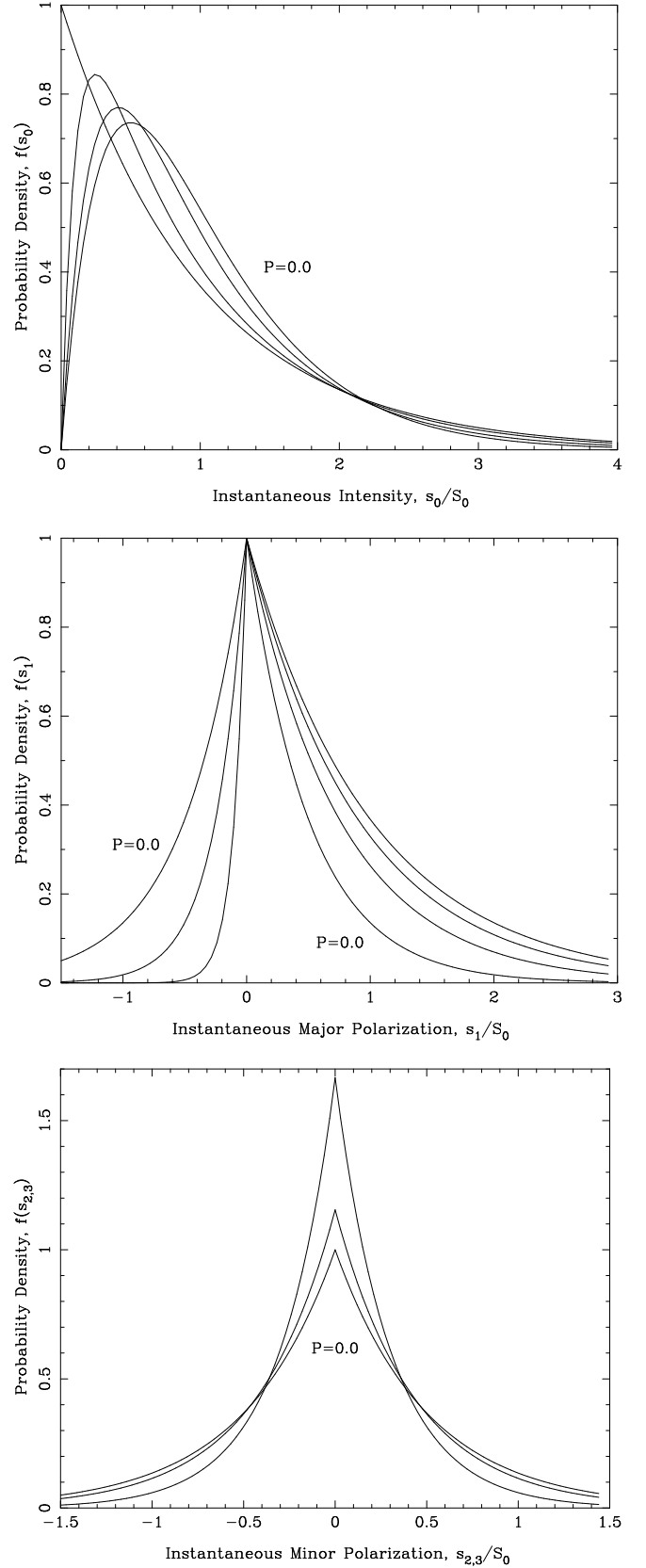

\begin{tabular}{l}
\includegraphics[angle=-90,width=86mm]{single_s0.eps} \\
\includegraphics[angle=-90,width=86mm]{single_s1.eps} \\
\includegraphics[angle=-90,width=86mm]{single_sm.eps}
\end{tabular}
\caption{\label{fig:single_marginal}
Probability densities of the instantaneous Stokes parameters as a
function of the degree of polarization, $P$.  From top to bottom are
the density functions of the total intensity, the major polarization,
and the minor polarizations.  In each panel, the unpolarized case is
labeled $P=0$ then followed by $P=0.5$, $P=0.8$ and, where applicable,
$P=1.0$.}
\end{figure}

\Eqns{single_intensity} through~(\ref{eqn:single_minor}) are plotted in
\Fig{single_marginal}.  The bottom two panels
illustrate the asymmetric, three-dimensional Laplace distribution of
the instantaneous polarization vector $\mbf{s}$; the top panel shows
the distribution of the magnitude of this vector $s_0=|\mbf{s}|$.
Qualitatively, instances of $\mbf{s}$ are distributed as a tapered
needle that points along the major polarization axis, with the
greatest density of instances in the head of the needle at the origin.
For unpolarized radiation, the distribution is spherically symmetric
and $f(s_1)=f(s_{2,3})$.  
For completely polarized radiation, $f(s_{2,3})=\delta(s_{2,3})$ and
$s_1=s_0$; that is, the distribution of $\mbf{s}$ becomes
one-dimensional and exponentially distributed along the positive $S_1$
axis.

%%%%%%%%%%%%%%%%%%%%%%%%%%%%%%%%%%%%%%%%%%%%%%%%%%%%%%%%%%%%%%%%%%%%%%
%%%%%%%%%%%%%%%%%%%%%%%%%%%%%%%%%%%%%%%%%%%%%%%%%%%%%%%%%%%%%%%%%%%%%%

\subsection{Local Means}
\label{sec:local}

\begin{figure}
\includegraphics[angle=-90,width=86mm]{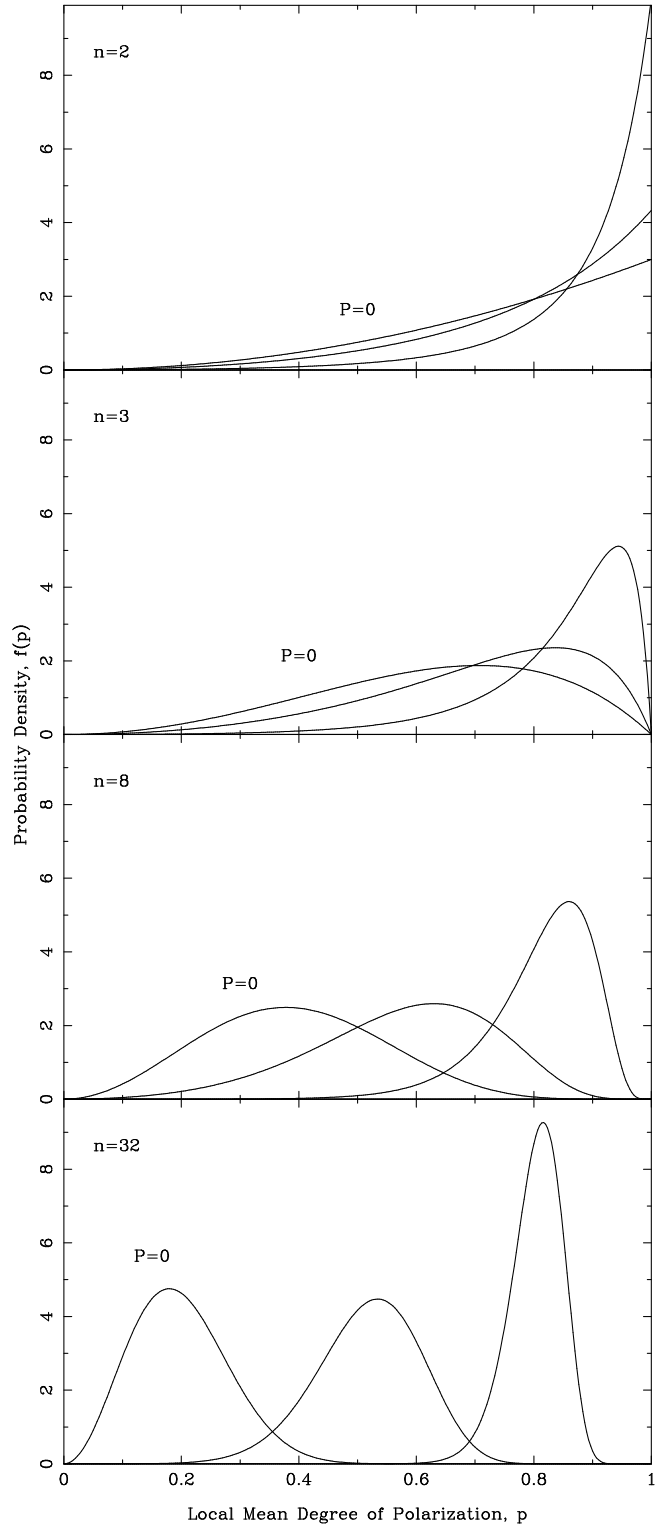}
\caption{
\label{fig:sample_p}
Probability densities of the local degree of polarization as function
of the intrinsic degree of polarization, $P$, and the number of
degrees of freedom, $n$. In each panel, the unpolarized case is
labeled $P=0$ then followed by $P=0.5$ and $P=0.8$. }
\end{figure}

\begin{figure}
\includegraphics[angle=-90,width=86mm]{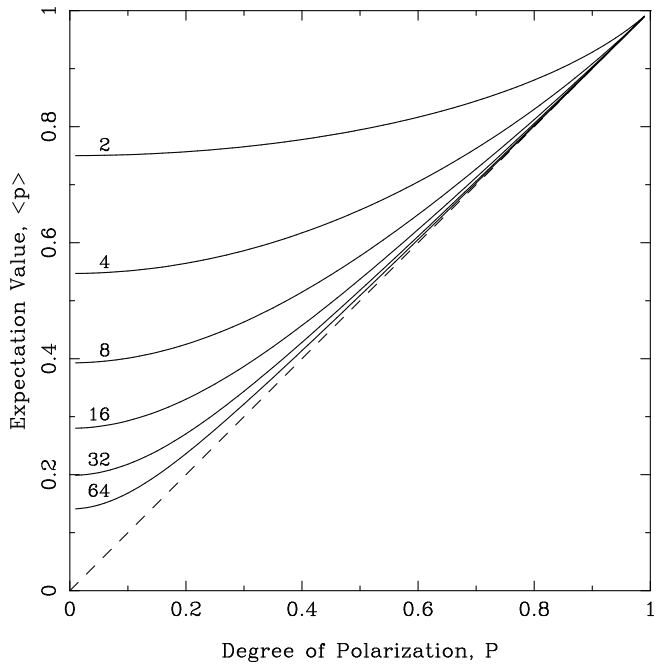}
\caption{
\label{fig:expected_p}
Expectation value of the local degree of polarization as a function of
the intrinsic degree of polarization.  The number of degrees of
freedom is printed above each curve and the dashed line indicates
$\langle p\rangle = P$. }
\end{figure}

\begin{figure}
\includegraphics[angle=-90,width=86mm]{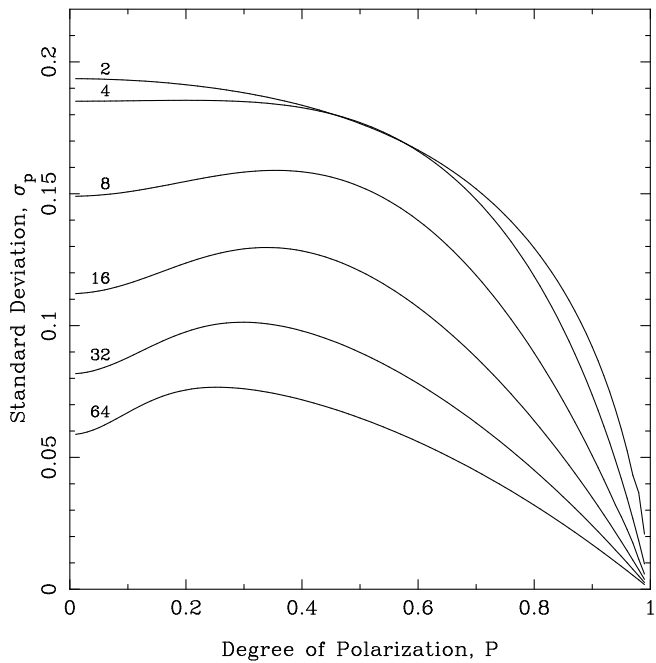}
\caption{
\label{fig:stddev_p}
Standard deviation of the local degree of polarization as a function of
the intrinsic degree of polarization.  The number of degrees of freedom
is printed above each curve. }
\end{figure}

The distribution of the local mean of $n>1$ independent and
identically distributed instances of the Stokes parameters $\mean[s]$
are derived from the distribution of the local mean coherency matrix
\begin{equation}
\label{eqn:local_rho}
\mean[\mbf\rho] = \frac{1}{n} \sum_{i=1}^n \mbf{e}_i \mbf{e}_i^\dagger
		= \frac{1}{2} \mean[s]_k \pauli{k},
\end{equation}
which has a complex Wishart distribution with $n$ degrees of freedom
\citep{wis28,goo63,tl96}; i.e.
\begin{equation}
f(\mean[\mbf\rho]) =
 { n^{2n} \left|\mean[\mbf\rho]\right|^{n-2}
   \over 
   \pi\Gamma(n)\Gamma(n-1)|\mbf{\rho}|^n }
 \exp \left( - n \trace \left[ \mbf{\rho}^{-1} \mean[\mbf\rho] \right] \right).
\end{equation}
The Wishart distribution is a multivariate generalization of the
$\chi^2$ distribution, and plays an important role in communications
engineering and information theory.
As in \Sec{single}, the Jacobian determinant
\begin{equation}
J(S_k,\mbf{\rho}) = \left|{\partial S_k\over\partial\mbf{\rho}}\right| = 8
\end{equation}
and \eqn{inverse} are used to arrive at the joint distribution
function of the local mean Stokes parameters,
\begin{equation}
f(\mean[s]) =
 { 2 n^{2n} \mean[s]^{2(n-2)}
   \over
   \pi\Gamma(n)\Gamma(n-1){S}^{2n} }
 \exp \left( - {2n\inner{S}{\mean[s]}\over \inv{S}}  \right).
\label{eqn:local}
\end{equation}
In \App{sample_p}, this joint distribution is used to derive the
probability density and the first two moments of the local mean degree
of polarization, $p=|\mean[\mbf{s}]|/\mean[s]_0$.
In \Fig{sample_p}, the distribution of $p$ is plotted as a function of
the intrinsic degree of polarization $P$ and the number of degrees of
freedom $n$.  Note that, when $P=1$, $f(p)=\delta(p-1)$; therefore, this
case is not shown.
The expected value of $p$ as a function $P$ and $n$ is plotted in
\Fig{expected_p}; the standard deviation $\sigma_p$ is similarly
plotted in \Fig{stddev_p}.

\Fig{expected_p} demonstrates that the self noise intrinsic to
a stationary stochastic source of polarized radiation induces a bias, 
$\langle p\rangle-P$, in the estimated degree of polarization.
The bias and standard deviation define the minimum sample
size required to estimate the degree of polarization to a certain
level of accuracy and precision.
Given the sample size and a measurement of $p$, the probability
density of $p$ (eqs.~[\ref{eqn:sample_p}] and ~[\ref{eqn:sample_p0}])
or the expectation value of $p$ (eq.~[\ref{eqn:sample_p_mean}]) could
be used to numerically estimate the intrinsic degree of polarization
using methods similar to those reviewed by \citet{ss85}.

%%%%%%%%%%%%%%%%%%%%%%%%%%%%%%%%%%%%%%%%%%%%%%%%%%%%%%%%%%%%%%%%%%%%%%
%%%%%%%%%%%%%%%%%%%%%%%%%%%%%%%%%%%%%%%%%%%%%%%%%%%%%%%%%%%%%%%%%%%%%%

\subsection{Ensemble Averages}
\label{sec:ensemble}
By the central limit theorem, at large $n$ the mean Stokes parameters
tend toward a multivariate normal distribution
\begin{equation}
\label{eqn:ensemble}
f(\smean[S]) = { 1 \over 4\pi^2 |{\bf C}|^{1/2} }
 \exp\left(-\frac{1}{2}\Delta S^{\rm T}{\bf C}^{-1}\Delta S\right),
\end{equation}
where $\Delta S=\smean[S]-S$ and ${\bf C}$ is the covariance matrix of
the Stokes parameters.
To derive the components of ${\bf C}$, consider a completely
unpolarized, dimensionless signal with unit intensity (that is, with
mean coherency matrix $\mbf{\rho}_u=\pauli{0}/2$ and mean Stokes
parameters $S_u=[1,0,0,0]$) and covariance matrix ${\bf
C}_u=\varsigma^2{\bf I}$, where $\varsigma^2$ is the dimensionless
variance of each Stokes parameter and ${\bf I}$ is the $4\times4$
identity matrix.
This unpolarized signal may be transformed into any partially
polarized signal with mean coherency matrix $\mbf\rho$ via a
congruence transformation
\begin{equation}
\mbf\rho = {\bf b} \mbf{\rho}_u {\bf b}^\dagger
= {\bf b b}^\dagger/2 = {\bf b}^2/2,
\end{equation}
where {\bf b} is the Hermitian square root of $2\mbf\rho$
\citep[e.g.][]{ham00} and the elements of $\mbf\rho$ have physical
dimensions such as flux density.
The covariance matrix of the Stokes parameters of the resulting
partially polarized signal is ${\bf{C}}={\bf{BC}}_u{\bf{B}}^T$, where
{\bf B} is the Mueller matrix of {\bf b}, defined by \eqn{Mueller}.
Noting that {\bf B} is symmetric, ${\bf{C}}=\varsigma^2{\bf{B}}^2$ is
simply a scalar multiple of the Mueller matrix of $\mbf\rho$; i.e.
\begin{equation}
\label{eqn:covariance}
C_{ij} = \varsigma^2 \left(2 S_iS_j - \eta_{ij} \inv{S} \right).
\end{equation}
This result is a generalization of equation (9) in \citet{bb92}, 
which derives from the complex Gaussian moment theorem.
In contrast, \eqn{covariance} requires no assumptions about the
distribution of the electric field.

The covariance matrix of the Stokes parameters has its simplest form
in the natural basis defined by the eigen decomposition of the
coherency matrix, where
\begin{equation}
\label{eqn:natural_covariance}
{\bf{C}}^\prime = \varsigma^2 \left(\begin{array}{cccc}
\norm{S}^2       &  2 S_0 |\mbf{S}| &  0 & 0 \\
2 S_0 |\mbf{S}|  & \norm{S}^2       &  0 & 0 \\
0 & 0 & S^2 & 0 \\
0 & 0 & 0 & S^2 \\
\end{array}\right).
\end{equation}
Comparison between the diagonal of ${\bf{C}}^\prime$ and the variances
of the distributions derived in \Sec{single} yields the relationship
between the dimensionless variance and the number of degrees of
freedom, $\varsigma^2=(2n)^{-1}$.
In the natural basis, it is also readily observed that the
polarization vector $\smean[\mbf{S}]$ is normally distributed as a
prolate spheroid with axial ratio, $\epsilon=((1+P^2)/(1-P^2))^{1/2}$.
The dimension of the major axis of the spheroid is equal to the
standard deviation of the total intensity and the dimension of the
minor axis goes to zero as the degree of polarization approaches
unity.
Furthermore, the multiple correlation between $\smean[S_0]$ and
$\smean[\mbf{S}]$ is
\begin{equation}
R = { 2 P \over 1 + P^2 }.
\label{eqn:multiple_correlation}
\end{equation}
The multiple correlation ranges from 0 for an unpolarized source to 1
for a completely polarized source.
It characterizes the correlation between total and polarized
intensities and expresses the fact that the stochastic Stokes
parameters cannot be treated in isolation.

%%%%%%%%%%%%%%%%%%%%%%%%%%%%%%%%%%%%%%%%%%%%%%%%%%%%%%%%%%%%%%%%%%%%%%
%%%%%%%%%%%%%%%%%%%%%%%%%%%%%%%%%%%%%%%%%%%%%%%%%%%%%%%%%%%%%%%%%%%%%%
%%%%%%%%%%%%%%%%%%%%%%%%%%%%%%%%%%%%%%%%%%%%%%%%%%%%%%%%%%%%%%%%%%%%%%

\section{Application to Radio Pulsar Astronomy}
\label{sec:application}

The previous section develops the four-dimensional statistics of the
Stokes parameters intrinsic to a single, stationary source of
stochastic electromagnetic radiation.
To apply these results to radio pulsar polarimetry, it is necessary to
consider various observed properties of pulsar signals, including
amplitude modulation, wave coherence, and the superposition of signals
from multiple sources, such as instrumental noise and orthogonally
polarized modes.
Emphasis is placed on the statistical degrees of freedom of the radiation;
in particular, the effects of amplitude modulation and wave coherence
on the {\it identically distributed} and {\it independent} conditions
of the central limit theorem are considered.
Attention is then focused on two areas of concern: randomly polarized
radiation and giant pulse polarimetry.

%%%%%%%%%%%%%%%%%%%%%%%%%%%%%%%%%%%%%%%%%%%%%%%%%%%%%%%%%%%%%%%%%%%%%%
%%%%%%%%%%%%%%%%%%%%%%%%%%%%%%%%%%%%%%%%%%%%%%%%%%%%%%%%%%%%%%%%%%%%%%

\subsection{Amplitude Modulation}
\label{sec:modulation}

Radio pulsar signals are well described as amplitude-modulated
noise \citep{ric75}.
The modulation index is generally defined as $\beta=\sigma_0/S_0$,
where $\sigma_0$ and $S_0$ are the standard deviation and mean of the
total intensity.
In single-pulse observations of radio pulsars, $\sigma_0$ is typically
corrected for the instrumental noise estimated from the off-pulse
baseline and the self noise is assumed to be negligible.
A value of $\beta$ greater than zero is then interpreted as evidence
of amplitude modulation.
However, for intense sources of radiation, the self noise of the
ensemble average total intensity, $\sigma_0=\varsigma\norm{S}$
(cf. eq.~[29] of Cordes 1976) must be taken into account.
That is, the self noise of the source induces a positive bias in the
modulation index.

Amplitude modulation modifies the statistics of all four Stokes
parameters.
For example, under scalar amplitude modulation, the covariance matrix
of the Stokes parameters becomes $\langle A^2 \rangle {\bf{C}}$, where
$A$ is the instantaneous intensity of the modulating function and
$\langle A^2 \rangle \ge \langle A \rangle^2 = 1$.
That is, amplitude modulation uniformly increases the covariances of
the Stokes parameters.
This can be interpreted as a reduction in the statistical degrees of
freedom of the signal, which becomes dominated by the fraction of
realizations that occur when $A \gg 1$.  That is, especially when the
modulations are deep, the samples are not identically distributed and
the central limit theorem does not trivially apply.

%%%%%%%%%%%%%%%%%%%%%%%%%%%%%%%%%%%%%%%%%%%%%%%%%%%%%%%%%%%%%%%%%%%%%%
%%%%%%%%%%%%%%%%%%%%%%%%%%%%%%%%%%%%%%%%%%%%%%%%%%%%%%%%%%%%%%%%%%%%%%

\subsection{Wave Coherence}
\label{sec:coherence}

The degrees of freedom of a stochastic process are also
reduced when the samples are not independent.
For plane-propagating electromagnetic radiation, statistical
dependence is manifest in various forms of wave coherence, including
that between the orthogonal components of the wave vector
(i.e. polarization) and that between instances of the field at different
coordinates (e.g. spectral, spatial, temporal).
In radio pulsar observations, wave coherence properties may be
modified by propagation in the pulsar magnetosphere and
the interstellar medium \cite[e.g.][]{lp00,mm98,mm00a}.

Given the coherence time $\tau_c$ \citep{man58,man59} of a
band-limited source of Gaussian noise, the effective number of degrees
of freedom is
\begin{equation}
N_{\rm eff} = { T \over \tau_c + \tau } \le {T\over\tau} = n,
\end{equation}
where $T$ is the integration interval, $\tau$ is the sampling
interval, and $n$ is the number of discrete time samples in the
interval $T$.
That is, owing to wave coherence, the signal could be encoded by
$N_{\rm eff}$ independent samples without any loss of information.

Substituting $n=N_{\rm eff}$ into \eqn{local} or
$\varsigma^2\propto1/N_{\rm eff}$ into \eqn{ensemble}, it is seen that
wave coherence inflates the four-dimensional volume occupied by the
joint distribution of the mean Stokes parameters.
With respect to the statistics of independent samples, this inflation
increases both the modulation index of the mean total intensity
and the eigenvalues of the covariance matrix of the mean Stokes
polarization vector \citep{es04,mck04}.
Referring to \Fig{expected_p}, it is readily observed that wave
coherence also increases the local mean degree of polarization.

If only the covariance matrix of the Stokes parameters is measured,
the effects of wave coherence are indistinguishable from those of
amplitude modulation.
In fact, the non-stationary statistics that arise from amplitude
modulation can be described by their spectral coherence properties
\cite[e.g.][]{bfs95}.
The various types of wave coherence may be differentiated via
auto-correlation and fluctuation spectral estimates
\citep[e.g.][]{cor76,es03,es04,jg04} that are outside the scope of the
current treatment.

%%%%%%%%%%%%%%%%%%%%%%%%%%%%%%%%%%%%%%%%%%%%%%%%%%%%%%%%%%%%%%%%%%%%%%
%%%%%%%%%%%%%%%%%%%%%%%%%%%%%%%%%%%%%%%%%%%%%%%%%%%%%%%%%%%%%%%%%%%%%%

\subsection{Incoherent Sum}
\label{sec:incoherent}

When one or more incoherent sources of radiation are added together,
the resulting covariance matrix of the total ensemble mean Stokes
parameters is the sum of the covariance matrices of the individual
sources.
For example, unpolarized noise adds a term to the diagonal of the
covariance matrix, such that
\[
C_{ij}^\prime = C_{ij} + \delta_{ij}\xi^2,
\]
thereby reducing both the ellipticity of the distribution of the
polarization vector and the multiple correlation between total and
polarized intensities.

The incoherent addition of signals with different polarizations,
especially the superposition of orthogonally polarized modes,
also decreases the degree of polarization of the mean Stokes parameters.
If the modes are covariant \citep[e.g.][]{ms98}, then the covariance
matrix of the sum includes cross-covariance terms.
For example, consider the incoherent sum of two sources described by
Stokes parameters $A$ and $B$.
If the intensities of the two modes are correlated, then the
resulting covariance matrix is given by
\begin{equation}
{\bf C} = {\bf C}_A + {\bf C}_B + \mbf{\Xi} + \mbf{\Xi}^T,
\label{eqn:covariant_sum}
\end{equation}
where ${\bf{C}}_A$ and ${\bf{C}}_B$ are defined as in \eqn{covariance}
and the cross-covariance matrix is
\begin{equation}
\Xi_{ij} = \varrho \varsigma_A \varsigma_B A_i B_j,
\label{eqn:cross_covariance}
\end{equation}
where $\varrho$ is the intensity correlation coefficient.
As shown in \App{covariant_opm}, the incoherent superposition of
covariant orthogonally polarized modes causes the variance of the
total intensity to increase while that of the major polarization
decreases.
(The major polarization is defined by the eigen decomposition and is
parallel to the major axis of the spheroidal distribution of $\mbf{S}$.)
Furthermore, it is shown that the four-dimensional covariance matrix
of the Stokes parameters can be used to derive the correlation
coefficient as well as the intensities and degrees of polarization of
superposed covariant orthogonal modes.

%%%%%%%%%%%%%%%%%%%%%%%%%%%%%%%%%%%%%%%%%%%%%%%%%%%%%%%%%%%%%%%%%%%%%%
%%%%%%%%%%%%%%%%%%%%%%%%%%%%%%%%%%%%%%%%%%%%%%%%%%%%%%%%%%%%%%%%%%%%%%

\subsection{Randomly Polarized Radiation}
\label{sec:rpr}

\citet{mck04} and \citet{es04} independently developed and applied
novel eigenvalue analyses of the three-dimensional covariance matrix
of the Stokes polarization vector $\mbf{S}$.
Each noted an apparent excess dispersion of $\mbf{S}$ and concluded
that it is due to the incoherent addition of randomly polarized
radiation intrinsic to the pulsar signal.
This hypothesis is based on the assumption that, apart from the
proposed randomly polarized component, the noise in each of the Stokes
parameters is purely instrumental, a premise that breaks down for
sources as bright as the pulsars studied in these experiments.

For example, \citet{mck04} analyzed observations of PSRs~B2020+28
and~B1929+10 that were recorded with the Arecibo 300~m antenna when
the forward gain was 8~K/Jy and the system temperature was 40~K
\citep{scr+84}.
Referring to the average pulse profiles presented in Figure 2 of
\citet{mck04}, the total intensity of PSR~B2020+28 peaks at 8~Jy
where the modulation index is $\sim 1.3$.
That is, the noise intrinsic to the pulsar signal exceeds the
system equivalent flux density (SEFD; $\sim5$~Jy) by as much as 100\%
\citep[cf. Figure 6 of][]{ms00}.
Similarly, the self noise of PSR~B1929+10 is as much as 50\% of the
SEFD.
In both cases, source-intrinsic noise statistics cannot be neglected.
To quantify the impact of self noise on single pulse polarimetry,
the results of \citet{mck04} are revisited.
Note that \citet{es04} focused on the non-orthogonal modes of
PSR~B0329+54 and reported neither the instrumental sensitivity nor the
statistics of the total intensity; therefore, this experiment is not
reviewed here.

Figures 3 and 4 of \citet{mck04} present two-dimensional projections
of the ellipsoidal distributions of $\mbf{S}$ at different pulse
phases, along with the dimensions of the major axis $a_1$ and minor
axes, $a_2$ and $a_3$, at each phase.  Panel a) in each of these plots
indicates the off-pulse, instrumental noise $\sigma_n$ in each
component of the Stokes polarization vector; for PSR~B2020+28,
$\sigma_n\simeq$ 0.1~Jy and, for PSR~B1929+10, $\sigma_n\simeq$
0.04~Jy.
Figures 2-4 are summarized in Table~1, which lists the pulse phase bin
number; the modulation index $\beta$; the total intensity $S_0$; and
the standard deviations of the total intensity $\sigma_0=\beta
S_0+\sigma_n$, the major polarization $\sigma_1=a_1$, and the minor
polarizations $\sigma_\perp=a_2\simeq a_3$.

\begin{table*}

\caption{Eigenvalue Analysis from \citet{mck04}}

\begin{center}
\begin{tabular}{llllll}

\tableline
\tableline

Bin & $\beta$ & $S_0$ (Jy)  & $\sigma_0$ & $\sigma_1$ & $\sigma_\perp$ \\

\tableline
\multicolumn{6}{c}{PSR B2020+28} \\
\tableline

61 & 0.45 & 5.5 & 2.58 & 1.14 & 0.59 \\
74 & 0.19 & 2.8 & 0.63 & 0.33 & 0.30 \\
91 & 0.56 & 3.1 & 1.84 & 0.91 & 0.71 \\

\tableline
\multicolumn{6}{c}{PSR B1929+10} \\
\tableline

90  & 0.55 & 0.68 & 0.41 & 0.32 & 0.09 \\
119 & 0.68 & 1.35 & 0.96 & 0.63 & 0.13 \\
125 & 0.65 & 1.00 & 0.69 & 0.46 & 0.10 \\

\tableline
\end{tabular}

\label{tab:mck04}
\end{center}
\end{table*}

As shown in \Sec{ensemble}, the standard deviation of the total
intensity and the major axis of the spheroidal distribution of the
polarization vector should be equal; however, for every phase bin
listed in Table 1, $\sigma_0>\sigma_1$.
That is, there is no excess dispersion of the polarization vector and
therefore no need for additional randomly polarized radiation;
rather, the excess dispersion of the total intensity requires
explanation.
Noting that scalar amplitude modulation and wave coherence uniformly
inflate the covariance matrix of the Stokes parameters, one possible
explanation for $\sigma_0>\sigma_1$ is the incoherent superposition of
covariant orthogonally polarized modes (see \App{covariant_opm}).
%%
%% REF 21
%%
The coefficient of correlation between the mode intensities $\varrho$
in \eqns{cov_opm_c00} and~(\ref{eqn:cov_opm_c11}) is equivalent to
$r_{12}$ in the analogous equations (5) and (6) of \citet{ms98}.
The last term in all four of these equations indicates that positive
correlation between the mode intensities increases the variance of the
total intensity ($\sigma_0^2=C_{00}+\sigma_n^2$) and decreases that of
the major polarization ($\sigma_1^2=C_{11}+\sigma_n^2$).
(Anti-correlated mode intensities have the opposite effect.)
The increased variance of the total intensity also explains why
orthogonally polarized modes typically coincide with increased
amplitude modulation \citep{mck04}.

Similarly, \eqn{cov_opm_c22} shows that the self noise of partially
polarized modes increases the dimensions of the minor axes
($a_2^2=C_{22}+\sigma_n^2$ and $a_3^2=C_{33}+\sigma_n^2$).  The
source-intrinsic contribution diminishes to zero only for 100\%
polarized modes.  For all of the phase bins listed in Table~1,
$\sigma_\perp>\sigma_n$, indicating that the modes are not completely
polarized, as has been previously assumed \citep{ms98,ms00}.

%%
%% REF 3
%%
\Eqns{cov_opm_c00} to~(\ref{eqn:cov_opm_c01}) and the discussion 
in \App{covariant_opm} motivate the development of a new technique for
producing mode-separated profiles.  From the four-dimensional
covariance matrix of the Stokes parameters, it is possible to derive
the mode intensity correlation coefficient as well as the intensities
and degrees of polarization of the modes.  These values and the mean
Stokes parameters, computed as a function of pulse phase, can be used
to decompose the pulse profile into the separate contributions of the
orthogonal modes.

%%%%%%%%%%%%%%%%%%%%%%%%%%%%%%%%%%%%%%%%%%%%%%%%%%%%%%%%%%%%%%%%%%%%%%
%%%%%%%%%%%%%%%%%%%%%%%%%%%%%%%%%%%%%%%%%%%%%%%%%%%%%%%%%%%%%%%%%%%%%%

\subsection{Giant Pulses}
\label{sec:giant}

The estimation of the degree of polarization of giant pulses is beset
by fundamental limitations.
On one hand, giant pulses remain unresolved at the highest time
resolutions achieved to date.
On the other hand, a sufficiently large number of independent samples
must be averaged before an accurate estimate of polarization is
possible.
Even if many time samples are averaged, those samples may be
correlated due to scattering on inhomogeneities in the interstellar
medium and/or the mean may be dominated by a single unresolved giant
nanopulse that greatly reduces the effective degrees of freedom.

For example, \citet{psk+06} presented a histogram of the degree of
circular polarization of giant pulses from the Crab pulsar observed
at 600~MHz.
The histogram is interpreted as evidence that each giant pulse is the
sum of $\sim100$ nanopulses, each with 100\% circular polarization.
However, referring to \Fig{stddev_p}, the width of the distribution is
also consistent with that of completely unpolarized radiation with
$\lesssim2$ degrees of freedom.
Although the data were averaged over 256 samples to 32~$\mu$s
resolution, the characteristic timescale of scattering in these
observations was estimated to be $45\pm5$~$\mu$s; therefore, the
effective number of degrees of freedom is of the order of unity.

Small-number statistics also limit the conclusions that can be
drawn from the correlations between giant pulse spectra presented
in Figure 12 of \citet{cbh+04}. 
The asymptotic correlation coefficient at small lag was interpreted as
evidence that nanopulses are highly polarized.
However, with only one estimate of the correlation coefficient at a
lag less than 0.1 seconds, the extrapolation to nanosecond resolution
is questionable.  Therefore, the constraint on the degree of
polarization at this timescale is of negligible significance.

%%%%%%%%%%%%%%%%%%%%%%%%%%%%%%%%%%%%%%%%%%%%%%%%%%%%%%%%%%%%%%%%%%%%%%
%%%%%%%%%%%%%%%%%%%%%%%%%%%%%%%%%%%%%%%%%%%%%%%%%%%%%%%%%%%%%%%%%%%%%%
%%%%%%%%%%%%%%%%%%%%%%%%%%%%%%%%%%%%%%%%%%%%%%%%%%%%%%%%%%%%%%%%%%%%%%

\section{Conclusion}
\label{sec:conclusion}

A four-dimensional statistical description of polarization is
presented that exploits the homomorphism between the Lorentz group and
the transformation properties of the Stokes parameters.
Within this framework, a generalized expression for the covariance
matrix of the Stokes parameters is developed and applied to the
analysis of single-pulse polarization.
The consideration of source-intrinsic noise renders randomly polarized
radiation unnecessary, explains the coincidence between increased
amplitude modulation and orthogonal mode fluctuations, and indicates
that orthogonally polarized modes are partially polarized.
Furthermore, the four-dimensional covariance matrix of the Stokes
parameters enables estimation of the mode intensities, degrees of
polarization, intensity correlation coefficient, and effective degrees
of freedom of covariant orthogonally polarized modes.
Measured as a function of pulse phase, these parameters may be used to
produce mode-separated profiles without any assumptions about the
intrinsic degree of mode polarization.

The formalism is also used to derive the first and second moments of
the degree of polarization as a function of the intrinsic degree of
polarization and the number of degrees of freedom of the stochastic
process.
These are used to demonstrate that giant pulse polarimetry is
fundamentally limited by systematic bias due to insufficient statistical
degrees of freedom.
The discussions of amplitude modulation in \Sec{modulation} and wave
coherence in \Sec{coherence} serve to illustrate the difficulties in
defining an appropriate average when the central limit theorem cannot
be trivially applied.
In this regard, the approach employed in this paper is complimentary
to previous analyses that make use of auto-correlation functions and
fluctuation spectra to separately measure the statistics of the total
and polarized intensities.
Useful new results may be derived by extending the techniques employed
in these works to include the cross-correlation terms that describe
the statistical dependences between the Stokes parameters.

\acknowledgments

I am grateful to J.P.~Macquart for fruitful discussions and K.~Stovall
for technical assistance with the computer algebra system.  The
insightful comments of the referee led to significant improvements to
the manuscript.  M.~Bailes, R.~Bhat, J.~Verbiest and M.~Walker also
provided helpful feedback on the text.

%%%%%%%%%%%%%%%%%%%%%%%%%%%%%%%%%%%%%%%%%%%%%%%%%%%%%%%%%%%%%%%%%%%%%%
%%%%%%%%%%%%%%%%%%%%%%%%%%%%%%%%%%%%%%%%%%%%%%%%%%%%%%%%%%%%%%%%%%%%%%
%%%%%%%%%%%%%%%%%%%%%%%%%%%%%%%%%%%%%%%%%%%%%%%%%%%%%%%%%%%%%%%%%%%%%%

\begin{appendix}

\section{Marginal Distributions of the Minor Stokes Parameters}
\label{app:marginal_minor}

\Eqns{single_intensity} and~(\ref{eqn:single_major}) could have
been also derived by first expressing \eqn{single_modes} in the natural
basis.  Here, it takes the simplified form,
\begin{equation}
f(|e_0|,|e_1|,\psi) = {1\over2\pi} f_0(|e_0|) f_1(|e_1|),
\label{eqn:single_modes_natural}
\end{equation}
where
\begin{equation}
f_m(|e_m|) = {2|e_m|\over\lambda_m} \exp \left( - |e_m|^2\over\lambda_m \right)
\end{equation}
are Rayleigh distributions.  Note that in the natural basis, $|e_0|$,
$|e_1|$, and $\psi$ are statistically independent and $\psi$ is
uniformly distributed.  The distributions of $s_0=|e_0|+|e_1|$ and
$s_1=|e_0|-|e_1|$ may then be obtained from the convolution and
cross-correlation, respectively, of $f_0(|e_0|)$ and $f_1(|e_1|)$.
Similarly, following \cite{fs81}, \eqn{single_modes_natural} is used
to compute the marginal distribution of $s_2=2|e_0||e_1|\cos\psi$.  In
the natural cylindrical coordinates defined in \Sec{single}, the
distribution of the radial dimension
$\Lambda=(s_2^2+s_3^2)^{-1/2}=2|e_0||e_1|$ is found by integrating
\eqn{single_modes_natural} over $\psi$, then transforming the
integrand to yield the intermediate result
\begin{equation}
f(\Lambda)
 = \int_0^\infty { f_0(|e_0|)f_1\left({\Lambda\over2|e_0|}\right) }
                  {1\over2|e_0|} d|e_0|
 = { \Lambda \over \lambda_0\lambda_1 }
      K_0 \left( {\Lambda \over \sqrt{\lambda_0\lambda_1} } \right),
\end{equation}
where $K_0$ is a modified Bessel function of the second kind.  As the
azimuthal dimension $\psi$ is uniformly distributed, the probability
density of $\cos\psi$ is
\[
f(\cos\psi)={1\over\pi\sqrt{1-\cos^2\psi}}.
\]
Combining with $f(\Lambda)$ and performing another integral
transformation yields
\begin{equation}
f(s_2)
 = \int_0^1 { f(\cos\psi)f\left({s_2\over\cos\psi}\right) }
                  {1\over\cos\psi} d\cos\psi
 = {1\over S} 
       \exp\left( -{ 2 s_2 \over S } \right)
\label{eqn:marginal_minor}
\end{equation}
for $s_2>0$, where $S$ is the Lorentz interval.  By symmetry, a
similar result is found for $s_2<0$.

\section{Distribution and Moments of the Local Mean Degree of Polarization}
\label{app:sample_p}

Conversion of \eqn{local} to spherical coordinates and integration
over all orientations of the polarization vector \mean[\mbf{s}] yields
the joint distribution of the local mean total and polarized
intensities,
\begin{equation}
f_n(\mean[s]_0,|\mean[\mbf{s}]|) = 
{ 4 n^{2n-1} |\mean[\mbf{s}]| \mean[s]^{2(n-2)}
  \over
  \Gamma(n)\Gamma(n-1) |\mbf{S}| S^{2(n-1)}
}
  \exp\left(-{2n \mean[s]_0 S_0 \over S^2}\right)
  \sinh\left({2n |\mean[\mbf{s}]| |\mbf{S}| \over S^2}\right)
\end{equation}
This joint density is defined on $0 \le |\mean[\mbf{s}]| \le
\mean[s]_0$ and is used to derive the distribution of the local mean
degree of polarization, $p=|\mean[\mbf{s}]|/\mean[s]_0$, as a function
of the intrinsic degree of polarization $P=|\mbf{S}|/S_0$ and
the number of samples averaged $n$:
\begin{equation}
\label{eqn:sample_p}
f(p) 
 = \int_0^\infty { \mean[s]_0 f_n(\mean[s]_0,p \mean[s]_0) } d\mean[s]_0 \\
 = { 2\Gamma(2n-1) \over \Gamma(n-1)\Gamma(n) }
   { (P^2 -1)^n p (p^2-1)^{n-2}
   \left( \left[ 1-Pp \right]^{1-2n} - \left[ 1+Pp \right]^{1-2n} \right)
   \over 2^{2n-1} P }.
\end{equation}
As $P\rightarrow0$,
\begin{equation}
\label{eqn:sample_p0}
f(p) 
 = { 2\Gamma(2n-1) \over \Gamma(n-1)\Gamma(n) }
   (-1)^n 2^{2-2n} (2n-1) p^2 (p^2 -1)^{n-2}.
\end{equation}
The distribution of the local mean degree of polarization is plotted
in \Fig{sample_p}; its first and second moments are
\begin{equation}
\label{eqn:sample_p_mean}
\langle p \rangle 
 = \int_0^1 p f(p) dp
 = (1-P^2)^n \Gamma(n+\frac{1}{2})
  \;{}_3\tilde{F}_2 \left(2,n,n+\frac{1}{2}; \frac{3}{2},n+1; P^2\right)
\end{equation}
and
\begin{equation}
\langle p^2 \rangle 
 = \int_0^1 p^2 f(p) dp
 = {3 (1-P^2)^n \over 1+2n }
  \;{}_3\tilde{F}_2 \left(\frac{5}{2},n,n+\frac{1}{2};
                      \frac{3}{2},n+\frac{3}{2}; P^2\right),
\end{equation}
where $_3\tilde{F}_2$ is a regularized generalized hypergeometric
function \citep{wol07}.  These moments are used to calculate the
variance of the local mean degree of polarization 
$\sigma_p^2=\langle p^2 \rangle - \langle p \rangle^2$.
The theoretical values of $\langle p \rangle$ and $\sigma_p$ are
plotted as a function of $n$ and $P$ in \Figs{expected_p}{stddev_p}.

%%%%%%%%%%%%%%%%%%%%%%%%%%%%%%%%%%%%%%%%%%%%%%%%%%%%%%%%%%%%%%%%%%%%%%
%%%%%%%%%%%%%%%%%%%%%%%%%%%%%%%%%%%%%%%%%%%%%%%%%%%%%%%%%%%%%%%%%%%%%%

\section{Covariant Orthogonal Partially Polarized Modes}
\label{app:covariant_opm}

Following the discussion in \Sec{incoherent}, the covariance matrix of
orthogonally polarized modes with correlated intensities is derived by
starting with \eqn{covariant_sum}, neglecting instrumental noise and
considering only the self noise of the modes.
If the modes $A$ and $B$ are orthogonally polarized, then
$\mbf{A \cdot B} = -|\mbf{A}||\mbf{B}|$; 
furthermore, if $A$ is the dominant mode, then $|\mbf{A}|>|\mbf{B}|$
and in the natural basis defined by $A$, the six non-zero elements of
the covariance matrix are
\begin{eqnarray}
\label{eqn:cov_opm_c00}
C_{00} & = & \varsigma_A^2 \norm{A}^2 + \varsigma_B^2 \norm{B}^2
	   + 2 \varrho \varsigma_A \varsigma_B A_0 B_0 \\
\label{eqn:cov_opm_c11}
C_{11} & = & \varsigma_A^2 \norm{A}^2 + \varsigma_B^2 \norm{B}^2
	   - 2 \varrho \varsigma_A \varsigma_B |\mbf{A}| |\mbf{B}| \\
\label{eqn:cov_opm_c22}
C_{22} = C_{33} & = & \varsigma_A^2 A^2 + \varsigma_B^2 B^2  \\
\label{eqn:cov_opm_c01}
C_{01} = C_{10} & = & 
        2 \varsigma_A^2 A_0|\mbf{A}| - 2 \varsigma_B^2 B_0 |\mbf{B}|
	+ \varrho \varsigma_A \varsigma_B ( B_0 |\mbf{A}| - A_0 |\mbf{B}| ).
\end{eqnarray}
Including $S_0=A_0+B_0$ and $|\mbf{S}|=|\mbf{A}|-|\mbf{B}|$,
there are a total of seven unknowns and six unique constraints.
However, if it is assumed that the modes have similar degrees of
freedom (i.e. $\varsigma_A\simeq\varsigma_B$), then the system can be
solved numerically; e.g. using the Newton-Raphson method \citep{ptvf92}.
In \citet{mck04}, $C_{01}$ is not measured; therefore, no derived
parameter estimates are currently presented.

\end{appendix}

\newpage

\begin{appendix}

  The following additional material was not submitted to The Astrophysical
  Journal and was not peer reviewed.  It is provided as further
  information for the interested reader. \\

\setcounter{section}{3}
\setcounter{equation}{13}

\section{Errors in this Manuscript}

The following errors in the published manuscript and in its published
Erratum have been discussed and addressed in \cite{vt17}.

\begin{enumerate}

\item In section 3.3, it is argued that Equation (28) is true regardless
  of the distribution of the electric field; however, it is true only in
  the case of jointly normally distributed electric field components.
  In general, the covariance matrix also depends on the Stokes cumulant
  as shown in Section 2.3 of \cite{vt17}.

\item In section 4.1, it is argued that amplitude modulation uniformly
  increases the covariances of the Stokes parameters (i.e. the
  covariance matrix is multiplied by a scalar greater than unity).  In
  fact, as shown in Section 4.2 of \cite{vt17}, scalar amplitude
  modulation increases the variance of the total intensity more than it
  increases the variances of the other three Stokes parameters.

\item In Section 4.3, it is incorrectly asserted that the covariance
  matrix of an incoherent sum of sources of radiation is simply the
  sum of the covariance matrices that describe the individual sources.
  Section 4.1 of \cite{vt17} presents the correct expression.
  
\end{enumerate}

%%%%%%%%%%%%%%%%%%%%%%%%%%%%%%%%%%%%%%%%%%%%%%%%%%%%%%%%%%%%%%%%%%%%%%%%%%%%%%
%%%%%%%%%%%%%%%%%%%%%%%%%%%%%%%%%%%%%%%%%%%%%%%%%%%%%%%%%%%%%%%%%%%%%%%%%%%%%%
%
% The Marginal Distributions
%
%%%%%%%%%%%%%%%%%%%%%%%%%%%%%%%%%%%%%%%%%%%%%%%%%%%%%%%%%%%%%%%%%%%%%%%%%%%%%%
%%%%%%%%%%%%%%%%%%%%%%%%%%%%%%%%%%%%%%%%%%%%%%%%%%%%%%%%%%%%%%%%%%%%%%%%%%%%%%

\section{Marginal Distributions}
\label{sec:marginal}

The marginal distributions of the Stokes parameters are computed in
the natural basis, where the axis of symmetry in cylindrical or
spherical coordinates is aligned with the major polarization, S1.

%%%%%%%%%%%%%%%%%%%%%%%%%%%%%%%%%%%%%%%%%%%%%%%%%%%%%%%%%%%%%%%%%%%%%%%%%%%%%%
%
% Single Samples: Total Intensity
%
%%%%%%%%%%%%%%%%%%%%%%%%%%%%%%%%%%%%%%%%%%%%%%%%%%%%%%%%%%%%%%%%%%%%%%%%%%%%%%

\subsection{Equation (19): Instantaneous Intensity}

To derive the marginal distribution of the instantaneous total intensity,
convert to spherical coordinates using the Jacobian determinant,
\begin{equation}
\left|{\partial\left(s_1,s_2,s_3\right)\over
       \partial\left(r,\theta,\phi\right)}\right| = r^2\cos\phi
\end{equation}
to arrive at the joint density,
\begin{equation}
f(r,\theta,\phi)= 
 {r\cos\phi \over \pi S^2}
 \exp\left[ -2\left( S_0r - |{\mbf S}|r\cos\phi\cos\theta \right)/ S^2\right].
\end{equation}
Note that $r=s_0$ and integrate over $\theta$ and $\phi$
(see {\tt Equation19.nb}) to yield
\begin{equation}
f(s_0) = {2\over|\mbf{S}|} \exp\left(-2{S_0\over S^2}s_0\right)
         \sinh\left(-2{|\mbf{S}|\over S^2}s_0\right).
\end{equation}
In terms of the eigenvalues,
\begin{equation}
f(s_0) = \left(\lambda_0-\lambda_1\right)^{-1}
         \left[\exp\left(-\lambda_0^{-1}s_0\right)
              -\exp\left(-\lambda_1^{-1}s_0\right)\right].
\end{equation}
This distribution has mean $S_0$ and variance $\norm{S}^2/2$.

%%%%%%%%%%%%%%%%%%%%%%%%%%%%%%%%%%%%%%%%%%%%%%%%%%%%%%%%%%%%%%%%%%%%%%%%%%%%%%
%
% Single Samples: Natural Polarization
%
%%%%%%%%%%%%%%%%%%%%%%%%%%%%%%%%%%%%%%%%%%%%%%%%%%%%%%%%%%%%%%%%%%%%%%%%%%%%%%

\subsection{Equation (20): Instantaneous Major Polarization}

To derive the marginal distribution of the major or natural
polarization, convert to cylindrical coordinates using the Jacobian
determinant,
\begin{equation}
\left|{\partial\left(s_1,s_2,s_3\right)\over
       \partial\left(t,\theta,s_1\right)}\right| = t
\end{equation}
to arrive at the joint density,
\begin{equation}
f(t,\theta,s_1)= 
 {t \over \pi S^2 s_0}
 \exp\left[ -2 \left( S_0s_0 - |{\mbf S}|s_1 \right)/ S^2\right].
\end{equation}
Note that $s_0^2=t^2+s_1^2$ and integrate over $\theta$ and $t$ 
(see {\tt Equation20.nb}) to yield
\begin{equation}
f(s_1) = {1\over S_0} \exp\left(-2{S_0|s_1|-|\mbf{S}|s_1 \over S^2}\right).
\end{equation}
In terms of the eigenvalues,
\begin{equation}
f(s_1)= \left(\lambda_0+\lambda_1\right)^{-1} 
\begin{cases}
        \exp\left(-\lambda_0^{-1}s_1\right) & s_1 > 0\\
        \exp\left(\lambda_1^{-1}s_1\right) & s_1 < 0\\
\end{cases}
\end{equation}
This distribution has mean $S_1$ and variance $\norm{S}^2/2$.

%%%%%%%%%%%%%%%%%%%%%%%%%%%%%%%%%%%%%%%%%%%%%%%%%%%%%%%%%%%%%%%%%%%%%%%%%%%%%%
%
% Single Samples: Minor Polarization
%
%%%%%%%%%%%%%%%%%%%%%%%%%%%%%%%%%%%%%%%%%%%%%%%%%%%%%%%%%%%%%%%%%%%%%%%%%%%%%%

\subsection{Equation (21): Instantaneous Minor Polarization}

This equation is derived in Appendix A of the paper.  The equations in
Appendix A are derived in {\tt AppendixA.nb}. \\ 

\noindent
Note that this distribution has mean $0$ and variance ${S}^2/2$.

%%%%%%%%%%%%%%%%%%%%%%%%%%%%%%%%%%%%%%%%%%%%%%%%%%%%%%%%%%%%%%%%%%%%%%%%%%%%%%
%
% Sample Means
%
%%%%%%%%%%%%%%%%%%%%%%%%%%%%%%%%%%%%%%%%%%%%%%%%%%%%%%%%%%%%%%%%%%%%%%%%%%%%%%

\subsection{Sample Means}

The distribution of the sample mean degree of polarization, $p$, as
a function of the number of samples, $n$, and the populations mean
degree of polarizaition, $P$, are derived in Appendix B of the paper.
The equations in Appendix B are derived in {\tt AppendixB.nb}.

%%%%%%%%%%%%%%%%%%%%%%%%%%%%%%%%%%%%%%%%%%%%%%%%%%%%%%%%%%%%%%%%%%%%%%%%%%%%%%
%%%%%%%%%%%%%%%%%%%%%%%%%%%%%%%%%%%%%%%%%%%%%%%%%%%%%%%%%%%%%%%%%%%%%%%%%%%%%%
%
% The Covariance Matrix of the Stokes parameters
%
%%%%%%%%%%%%%%%%%%%%%%%%%%%%%%%%%%%%%%%%%%%%%%%%%%%%%%%%%%%%%%%%%%%%%%%%%%%%%%
%%%%%%%%%%%%%%%%%%%%%%%%%%%%%%%%%%%%%%%%%%%%%%%%%%%%%%%%%%%%%%%%%%%%%%%%%%%%%%

\section{Equation (28): Stokes Covariance Matrix}
\label{sec:covariance}

%%%%%%%%%%%%%%%%%%%%%%%%%%%%%%%%%%%%%%%%%%%%%%%%%%%%%%%%%%%%%%%%%%%%%%%%%%%%%%
%
% The short way
%
%%%%%%%%%%%%%%%%%%%%%%%%%%%%%%%%%%%%%%%%%%%%%%%%%%%%%%%%%%%%%%%%%%%%%%%%%%%%%%

\subsection{The short way}

For the simplest derivation of the covariance matrix, note the following:

\begin{itemize}

\item A unitary transformation does not alter the degree of
polarization; therefore, {\bf b} must be Hermitian and ${\bf
b}^2=2\mbf{\rho}$.

\item The Mueller matrix {\bf B} of the Jones matrix {\bf b} is a
Lorentz transformation, which is symmetric; therefore ${\bf C}={\bf B}^2$.

\item The Mueller matrix ${\bf B}^2$ corresponds to the Jones matrix
$2\mbf{\rho}$, which is also Hermitian; therefore, {\bf C} is also 
a Lorentz transformation.

\end{itemize}

\noindent
Refer to Equation (12) of Britton (2000) for the axis-angle
parameterization of a Lorentz transformation, and substitute the Jones
matrix
\[
\boost = 2\mbf{\rho} = S_i\pauli{i}.
\]
Here, it is useful to note that
\begin{eqnarray}
\cosh2\beta = & 2\cosh^2\beta-1 & = 2S_0^2 -1 \nonumber \\
\sinh2\beta \,m_k = & 2\cosh\beta \sinh\beta \, m_k & = 2S_0S_k \nonumber
\end{eqnarray}

%%%%%%%%%%%%%%%%%%%%%%%%%%%%%%%%%%%%%%%%%%%%%%%%%%%%%%%%%%%%%%%%%%%%%%%%%%%%%%
%
% The long way
%
%%%%%%%%%%%%%%%%%%%%%%%%%%%%%%%%%%%%%%%%%%%%%%%%%%%%%%%%%%%%%%%%%%%%%%%%%%%%%%

\subsection{The long way}

Alternatively, you can start with the definition of the
covariance matrix, ${\bf C}={\bf B\,B}^{\rm T}$, or
\begin{equation}
C_{ij} = B_i^k B_j^k 
= {1\over4}
  \trace\left(\pauli{i}\,{\bf b}\,\pauli{k}\,{\bf b}^\dagger \right)
  \trace\left(\pauli{j}\,{\bf b}\,\pauli{k}\,{\bf b}^\dagger \right)
\end{equation}
and note that the trace is

\begin{itemize}
\item a scalar: $a=\trace({\bf A})$;
\item linear: $a\trace({\bf B})=\trace(a{\bf B})$;
\item commutative: $\trace({\bf AB}) = \trace({\bf BA})$; and
\item a projection operator: ${\bf A}=\trace({\bf A}\pauli{i})\pauli{i}/2$.
\end{itemize}
Therefore,
\begin{eqnarray}
C_{ij}
& = & {1\over4}
 \trace\left[
            \trace\left({\bf b}^\dagger\,\pauli{i}\,{\bf b}\,\pauli{k} \right)
            \pauli{k}\,{\bf b}^\dagger\,\pauli{j}\,{\bf b} 
      \right] \\
& = & {1\over2} 
    \trace\left({\bf b}^\dagger\,\pauli{i}\,{\bf b}\,
                {\bf b}^\dagger\,\pauli{j}\,{\bf b}\right) \\
& = & 2 \trace\left(\pauli{i}\,\mbf{\rho}\,\pauli{j}\,\mbf{\rho}\right),
\end{eqnarray}
which is the Mueller matrix of $2\mbf\rho$.  Again, it is possible to
refer to Equation (12) of Britton (2000) or use the trace of the
anticommutator,
\begin{equation}
\trace(\acomm{\bf A}{\bf B})= 2\trace({\bf AB})
\end{equation}
to show that
\begin{equation}
C_{ij}
= 2 \trace\left(\acomm{\pauli{i}\,\mbf{\rho}}{\pauli{j}\,\mbf{\rho}}\right)
= {1\over2} S_k S_l 
  \trace\left(\acomm{\pauli{i}\,\pauli{k}}{\pauli{j}\,\pauli{l}}\right).
\end{equation}
The anticommutator of the Pauli matrices,
\begin{equation}
\acomm{\pauli{i}}{\pauli{j}} = 2 \delta_{ij}\pauli{0}.
\end{equation}
Therefore, only 4 of the 16 terms in the above double sum do not
vanish.  When $i = j$, the four $k=l$ terms remain.  If $i=j=0$,
the result is simply twice the Frobenius norm, 
\begin{equation}
C_{00} = 2 \trace (\mbf{\rho}^\dagger\mbf{\rho})
       = S_k S_k = 2 S_0^2 - |S|
\end{equation}
For $i = j > 0$, two terms are positive and two are negative:

\begin{enumerate}

\item $k=l=0$: $\trace (\acomm{\pauli{i}}{\pauli{i}}) = 4$;

\item $k=l=i=j$: $\trace (\acomm{\pauli{0}}{\pauli{0}}) = 4$;

\item[3$-$4.] otherwise: $\trace (\acomm{i\epsilon_{ik\alpha}\pauli{\alpha}}{i\epsilon_{ik\alpha}\pauli{\alpha}}) = 4i^2=-4$. 

\end{enumerate}
Therefore,
\begin{equation}
C_{\alpha\alpha} = S_0^2 + S_\alpha^2 - S_\beta^2 - S_\gamma^2
                 = 2 S_\alpha^2 + |S|,
\end{equation}
where $\alpha$, $\beta$, and $\gamma$ are all greater than zero and unequal.
Combining the two equations yields
\begin{equation}
C_{ii} = 2 S_i^2 -\eta_{ii} |S|
\end{equation}
(no summation is implied by repeated indeces).  When $i \neq j$, the
four terms that remain are
\begin{enumerate}
\item $k=i$ and $l=j$: the arguments to the anticommutator are the identity
matrix and the trace is 4;
\item $k=j$ and $l=i$: if either $i$ or $j$ is zero, then the arguments to
the anticommutator are the other
matrix and the trace is 4; otherwise,
$\pauli{i}\pauli{j}=-\pauli{j}\pauli{i}=i\epsilon_{ijk}\pauli{k}$ and
the trace is $-4i^2=4$.
\item $k=0$ and $\epsilon_{jli}\neq0$:
 $\trace (\acomm{\pauli{i}}{\pm i\pauli{i}}) = \pm 4i$;
\item $l=0$ and $\epsilon_{ikj}\neq0$:
 $\trace (\acomm{\mp i\pauli{j}}{\pauli{j}}) = \mp 4i$.
\end{enumerate}
The last two terms cancel eachother, and $C_{ij}=2S_iS_j$.

%%%%%%%%%%%%%%%%%%%%%%%%%%%%%%%%%%%%%%%%%%%%%%%%%%%%%%%%%%%%%%%%%%%%%%%%%%%%%%
%
% The Determinant
%
%%%%%%%%%%%%%%%%%%%%%%%%%%%%%%%%%%%%%%%%%%%%%%%%%%%%%%%%%%%%%%%%%%%%%%%%%%%%%%

\subsection{In the Eigen Basis}
\label{sec:eigen_basis}

In the eigen basis, the covariance matrix
\begin{equation}
{\bf{C}} = \left(\begin{array}{cccc}
S_0^2 + |\mbf{S}|^2  &  2 S_0 |\mbf{S}|     &  0 & 0 \\
2 S_0 |\mbf{S}|      & S_0^2 + |\mbf{S}|^2  &  0 & 0 \\
0 & 0 & S^2 & 0 \\
0 & 0 & 0 & S^2 \\
\end{array}\right)
\end{equation}
from which it is trivial to derive the determinant by Laplacian expansion,
\begin{equation}
|{\bf{C}}| = \left( S_0^2 + |\mbf{S}|^2 \right)^2 S^4 
           - \left( 2 S_0 |\mbf{S}| \right)^2 S^4 
           = S^8
\end{equation}

\end{appendix}


\begin{thebibliography}{}
\expandafter\ifx\csname natexlab\endcsname\relax\def\natexlab#1{#1}\fi

\bibitem[{Backer \& Rankin(1980)}]{br80}
Backer, D.~C., \& Rankin, J.~M. 1980, ApJS, 42, 143

\bibitem[{{Barakat}(1963)}]{bar63}
{Barakat}, R. 1963, J. Opt. Soc. Am., 53, 317

\bibitem[{Barakat(1987)}]{bar87}
Barakat, R. 1987, J. Opt. Soc. Am. A, 4, 1256

\bibitem[{{Bertolotti} {et~al.}(1995){Bertolotti}, {Ferrari}, \&
  {Sereda}}]{bfs95}
{Bertolotti}, M., {Ferrari}, A., \& {Sereda}, L. 1995, J. Opt. Soc. Am. B, 12,
  341

\bibitem[{Born \& Wolf(1980)}]{bw80}
Born, M., \& Wolf, E. 1980, Principles of optics: electromagnetic theory of
  propagation, interference and diffraction of light (New York: Pergamon)

\bibitem[{Britton(2000)}]{bri00}
Britton, M.~C. 2000, ApJ, 532, 1240

\bibitem[{Brosseau \& Barakat(1992)}]{bb92}
Brosseau, C., \& Barakat, R. 1992, Opt. Comm., 91, 408

\bibitem[{Cairns {et~al.}(2001)Cairns, Johnston, \& Das}]{cjd01}
Cairns, I.~H., Johnston, S., \& Das, P. 2001, ApJ, 563, L65

\bibitem[{Cloude(1986)}]{clo86}
Cloude, S. 1986, Optik, 75, 26

\bibitem[{Cognard {et~al.}(1996)Cognard, Shrauner, Taylor, \&
  Thorsett}]{cstt96}
Cognard, I., Shrauner, J.~A., Taylor, J.~H., \& Thorsett, S.~E. 1996, ApJ, 457,
  L81

\bibitem[{{Cordes}(1976)}]{cor76}
{Cordes}, J.~M. 1976, ApJ, 208, 944

\bibitem[{{Cordes} {et~al.}(2004){Cordes}, {Bhat}, {Hankins}, {McLaughlin}, \&
  {Kern}}]{cbh+04}
{Cordes}, J.~M., {Bhat}, N.~D.~R., {Hankins}, T.~H., {McLaughlin}, M.~A., \&
  {Kern}, J. 2004, ApJ, 612, 375

\bibitem[{Cordes \& Hankins(1977)}]{ch77}
Cordes, J.~M., \& Hankins, T.~H. 1977, ApJ, 218, 484

\bibitem[{{Edwards} \& {Stappers}(2003)}]{es03}
{Edwards}, R.~T., \& {Stappers}, B.~W. 2003, A\&A, 407, 273

\bibitem[{{Edwards} \& {Stappers}(2004)}]{es04}
---. 2004, A\&A, 421, 681

\bibitem[{Eliyahu(1994)}]{eli94}
Eliyahu, D. 1994, Phys. Rev. E, 50, 2381

\bibitem[{Fercher \& Steeger(1981)}]{fs81}
Fercher, A.~F., \& Steeger, P.~F. 1981, Optica Acta, 28, 443

\bibitem[{Goodman(1963)}]{goo63}
Goodman, N.~R. 1963, Ann. of Math. Stat., 34, 152

\bibitem[{{Hamaker}(2000)}]{ham00}
{Hamaker}, J.~P. 2000, A\&AS, 143, 515

\bibitem[{{Hankins} {et~al.}(2003){Hankins}, {Kern}, {Weatherall}, \&
  {Eilek}}]{hkwe03}
{Hankins}, T.~H., {Kern}, J.~S., {Weatherall}, J.~C., \& {Eilek}, J.~A. 2003,
  Nature, 422, 141

\bibitem[{Heiles {et~al.}(1970)Heiles, Campbell, \& Rankin}]{hcr70}
Heiles, C., Campbell, D.~B., \& Rankin, J.~M. 1970, Nature, 226, 529

\bibitem[{{Jenet} \& {Gil}(2004)}]{jg04}
{Jenet}, F.~A., \& {Gil}, J. 2004, ApJ, 602, L89

\bibitem[{Lyutikov \& Parikh(2000)}]{lp00}
Lyutikov, M., \& Parikh, A. 2000, ApJ, 541, 1016

\bibitem[{{Macquart} \& {Melrose}(2000)}]{mm00a}
{Macquart}, J.-P., \& {Melrose}, D.~B. 2000, 62, 4177

\bibitem[{Manchester {et~al.}(1975)Manchester, Taylor, \& Huguenin}]{mth75}
Manchester, R.~N., Taylor, J.~H., \& Huguenin, G.~R. 1975, ApJ, 196, 83

\bibitem[{Mandel(1958)}]{man58}
Mandel, L. 1958, Proc.\ Phys.\ Soc., 72, 1037

\bibitem[{Mandel(1959)}]{man59}
---. 1959, Proc.\ Phys.\ Soc., 74, 233

\bibitem[{Mandel(1963)}]{man63}
---. 1963, Proc.\ Phys.\ Soc., 81, 1104

\bibitem[{McKinnon \& Stinebring(1998)}]{ms98}
McKinnon, M., \& Stinebring, D. 1998, ApJ, 502, 883

\bibitem[{{McKinnon}(2002)}]{mck02}
{McKinnon}, M.~M. 2002, ApJ, 568, 302

\bibitem[{{McKinnon}(2003{\natexlab{a}})}]{mck03a}
---. 2003{\natexlab{a}}, ApJ, 590, 1026

\bibitem[{{McKinnon}(2003{\natexlab{b}})}]{mck03b}
---. 2003{\natexlab{b}}, ApJS, 148, 519

\bibitem[{{McKinnon}(2004)}]{mck04}
---. 2004, ApJ, 606, 1154

\bibitem[{{McKinnon}(2006)}]{mck06}
---. 2006, ApJ, 645, 551

\bibitem[{{McKinnon} \& {Stinebring}(2000)}]{ms00}
{McKinnon}, M.~M., \& {Stinebring}, D.~R. 2000, ApJ, 529, 435

\bibitem[{Melrose \& Macquart(1998)}]{mm98}
Melrose, D.~B., \& Macquart, J.-P. 1998, ApJ, 505, 921

\bibitem[{{Popov} {et~al.}(2006){Popov}, {Soglasnov}, {Kondrat'ev}, {Kostyuk},
  {Ilyasov}, \& {Oreshko}}]{psk+06}
{Popov}, M., {Soglasnov}, V., {Kondrat'ev}, V., {et~al.} 2006, Astronomy
  Letters, 50, 55

\bibitem[{Press {et~al.}(1992)Press, Teukolsky, Vetterling, \&
  Flannery}]{ptvf92}
Press, W.~H., Teukolsky, S.~A., Vetterling, W.~T., \& Flannery, B.~P. 1992,
  Numerical Recipes: {T}he Art of Scientific Computing, 2$^{nd}$ edition
  (Cambridge: Cambridge University Press)

\bibitem[{{Rickett}(1975)}]{ric75}
{Rickett}, B.~J. 1975, ApJ, 197, 185

\bibitem[{{Simmons} \& {Stewart}(1985)}]{ss85}
{Simmons}, J.~F.~L., \& {Stewart}, B.~G. 1985, A\&A, 142, 100

\bibitem[{{Steeger} {et~al.}(1984){Steeger}, {Asakura}, {Zocha}, \&
  {Fercher}}]{sazf84}
{Steeger}, P.~F., {Asakura}, T., {Zocha}, K., \& {Fercher}, A.~F. 1984, J. Opt.
  Soc. Am. A, 1, 677

\bibitem[{Stinebring {et~al.}(1984)Stinebring, Cordes, Rankin, Weisberg, \&
  Boriakoff}]{scr+84}
Stinebring, D.~R., Cordes, J.~M., Rankin, J.~M., Weisberg, J.~M., \& Boriakoff,
  V. 1984, ApJS, 55, 247

\bibitem[{Taylor {et~al.}(1971)Taylor, Huguenin, Hirsch, \&
  Manchester}]{thhm71}
Taylor, J.~H., Huguenin, G.~R., Hirsch, R.~M., \& Manchester, R.~N. 1971,
  Astrophys. Lett., 9, 205

\bibitem[{Touzi \& Lopes(1996)}]{tl96}
Touzi, R., \& Lopes, A. 1996, I. E. E. E. Trans. Geoscience and Remote Sensing,
  34, 519

\bibitem[{{van Straten} \& {Tiburzi}(2017)}]{vt17}
{van Straten}, W., \& {Tiburzi}, C. 2017, ApJ, 835, 293

\bibitem[{Wishart(1928)}]{wis28}
Wishart, J. 1928, Biometrika, 20A, 32

\bibitem[{Wolleben(2007)}]{wol07}
Wolleben, M. 2007, ApJ, 664, 349

\end{thebibliography}
\end{document}